\documentclass[conference]{IEEEtran}
\ifCLASSINFOpdf
\else
   \usepackage[dvips]{graphicx}
\fi
\usepackage[cmex10]{amsmath}
\usepackage{bm}
\begin{document}
%
\title{Joint Viterbi Decoding and Decision Feedback Equalization for Monobit Digital Receivers}

\author{
\IEEEauthorblockN{Xin Li, Huarui Yin, Zhiyong Wang}
\IEEEauthorblockA{Department of Electronic Engineering and Information Science\\
University of Science \& Technology of China\\Hefei, 230027, China\\
Email: yhr@ustc.edu.cn, \{rich1989, wzy2\}@mail.ustc.edu.cn}
\and
\IEEEauthorblockN{Zhengdao Wang}
\IEEEauthorblockA{Department of Electrical and Computer Engineering \\Iowa State University\\Ames, IA, 50011, USA\\ Email: zhengdao@iastate.edu}

}



%


\maketitle

\begin{abstract}

In ultra-wideband (UWB) communication systems with impulse radio (IR) modulation, the bandwidth is usually 1GHz or more. To process the received signal digitally, high sampling rate analog-digital-converters (ADC) are required. Due to the high complexity and large power consumption, monobit ADC is appropriate. The optimal monobit receiver has been derived. But it is not efficient to combat intersymbol interference (ISI). Decision feedback equalization (DFE) is an effect way dealing with ISI. In this paper, we proposed a algorithm that combines Viterbi decoding and DFE together for monobit receivers. In this way, we suppress the impact of ISI effectively, thus improving the bit error rate (BER) performance. By state expansion, we achieve better performance. The simulation results show that the algorithm has about 1dB SNR gain compared to separate demodulation and decoding method and 1dB loss compared to the BER performance in the channel without ISI. Compare to the full resolution detection in fading channel without ISI, it has 3dB SNR loss after state expansion.

\end{abstract}


%
\IEEEpeerreviewmaketitle

\section{Introduction}

\begin{figure*}[!t]
\centering
\includegraphics[width=7in]{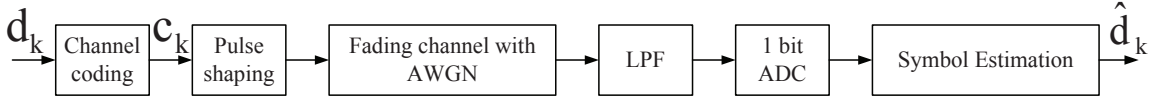}
\caption{System Blockdiagram}
\label{sysblock}
\normalsize
\vspace*{4pt}
\end{figure*}

Impulse radio ultra-wideband (IR-UWB) systems utilize the short pulse with a low duty cycle to carry information \cite{UWBintroduction}. The short duration of the pulse enables high data rate but occupies large bandwidth, usually 1 GHz or more. In communication systems, digital approaches can provide attractive flexibility in receiver signal processing. In order to process the received signal digitally, the analog-digital-converter (ADC) is required. Due to the ADC resolution and the limitation of its power consumption, it is appropriate to use low-resolution ADC.  Monobit ADC has a simple structure and can be realized by a fast comparator. Tens of Giga samples per second (Gsps) could be reached. Monobit ADCs have low power consumption and low cost. 

Monobit digital receivers for IR-UWB have been developed, see e.g. \cite{0102}. Optimal monobit receiver under Nyquist sampling rate has been proposed in \cite{Optimal}. The optimal receiver turns out to be a linear combiner. Suboptimal monobit receiver was also proposed.

In indoor channel, due to the multipath propagation, the communication system suffers from the effect of intersymbol interference (ISI). The receiver has to detect symbols in the ISI. The optimal monobit receiver mentioned above was derived under the assumption that maximum channel delay is significantly smaller than symbol duration. Thus it is may not perform well when the delay spread is large. 

There are plenty of ways to cope with ISI \cite{0103}. Current strategies that suppress ISI mainly include the following: the equalizer based on minimum mean square error (MMSE) rule, which is introduced and analyzed in \cite{MMSE}, the Zero Forcing (ZF) Equalizer in the Bell System Technical Journal in 1965 \cite{ZF} and the decision-feedback equalizer (DFE) which cancels the ISI by feeding back the decided results \cite{DFE}. Among them, DFE is relatively a simple approach. At high SNR, this method makes reliable decisions. However, at low SNR, error propagation is a severe problem. In practical communication systems, channel coding is a critical component. Under full resolution (FR) sampling, a joint coding and DFE method has been proposed in \cite{joint}. This method uses a combination of Viterbi soft decisions and delayed decisions to minimize bit error rate (BER). Unfortunately, in UWB systems, as a result of the short symbol duration, full resolution received waveform is difficult to obtain. For monobit receivers, no algorithms have been developed to combat heavy ISI as far as we know.

In this paper, we propose a joint Viterbi decoding and DFE algorithm for monobit receivers. This algorithm provides an efficient way for monobit receivers dealing with ISI in indoor fading channel. The channel has one line-of-sight (LOS) path and many non-line-of-sight (NLOS) paths. The NLOS paths appear in a relatively long time after the LOS arose. In this algorithm, for every arrived state in the code trellis, we construct a full-resolution reference waveform to cancel the overlapped ISI, then we use derived monobit optimal approach to compute the path metric \cite{Optimal}. The path having the maximum likelihood probability metric remains and is set as the current surviving path. We will derive the likelihood probability for each state. State expansion is also proposed to improve the BER performance. The simulation results show that compared to the fading channel without ISI, the receiver suffered 2dB SNR loss from ISI. After state expansion, the loss reduces to 1dB. Compared to FR detection in the fading channel without ISI, the receiver has 3dB SNR loss. Compared to separate demodulation and decoding, the joint decoding and DFE algorithm has 1dB SNR gain.

The rest of the paper is organized as follows: In Section \uppercase\expandafter{\romannumeral2}, we describe the system model. Section \uppercase\expandafter{\romannumeral3} describes the algorithm we proposed. We give an example and discuss state expansion method. We can use this approach get better BER performance but the complexity increases. Simulation results are provided in Section \uppercase\expandafter{\romannumeral4}. Section \uppercase\expandafter{\romannumeral5} is the conclusion.

\section{System Model}

The block diagram of joint Viterbi decoding and DFE monobit digital receivers is depicted in Fig. \ref{sysblock}. The baseband received signal is first filtered by an ideal low pass filter (LPF). The bandwidth of the filter is $B$. Then the received signal will be sampled at Nyquist sampling rate $T = 1/2B$ and quantized to one bit resolution. The digitized signal is processed by a digital signal processing (DSP) unit for decoding and symbol estimation.

In this paper, the transmitted information are binary symbols. We assume every data block has $U$ binary symbols. $d_u \in \{ +1,-1\}$ is the $u$th symbol which is equally likely to be $\pm 1$. The transmitted information bits are first encoded by a convolutional encoder. The rate of the encoder is $R = 1/2$. We have the code symbols $c_k,k\in[0,2U-1]$. Let the vector $\bm{c_m}=[~c_0,~c_1,~...,~c_m~]$ denote the first $m(m \le 2U-1)$ code bits. We assume the modulation type is binary pulse amplitude modulation (PAM). The transmitted signal can be written as
\begin{equation}
s(t)=\sum_{k=0}^{2U-1}c_k p_\text{tr}(t-kT_s)
\end{equation}
where $p_\text{tr}$ is the shaping pulse, and $T_s$ is symbol duration.

The wireless channel can be modeled as a linear time-invariant (LTI) system with a finite impulse response $h(t)=\sum_{l=0}^{L-1}\alpha_l\delta(t-\tau_l)$. In the case of wireless time-varying channel, we assume that within the coherent interval the
channel can be modeled as time-invariant. The multipath delay spread of the indoor channel is $T_\text{max}$. The ISI appears if $T_\text{max} > T_s $. 

The received signal can be written as $r(t)=s(t)\star h(t)+n(t)$, where $\star$ denotes convolution. $n\left(t\right)$ is AWGN with double-sided power spectral density $N_0/2$. The system function of the ideal LPF is 

\begin{displaymath}
H_\text{lp}(\omega) = \left\{ \begin{array}{ll}
 1/\sqrt{N_0B}  & \textrm{$|\omega|\le B$}\\
0 & \textrm{others}\\
\end{array} \right.
\end{displaymath}
The variance of noise will be normalized to one after filtered by the LPF. 
The filtered $r(t)$ can be expressed as 
\begin{equation}
r(t)=\sum_{k=0}^{2U-1}c_k p_\text{ref}(t-kT_s)+n'(t)
\end{equation}
where $p_\text{ref}(t) = p_\text{tr}(t) \star h(t) \star h_\text{lp}(t)$, $n'(t) = n(t) \star h_\text{lp}(t)$. $h_\text{lp}(t)$ is the impulse response of the LPF.

Define sampling period $T = 1/(2B) = T_s/N$, which means for every pulse in the duration $T_s$, $N$ samples are generated. The received signal $r(t)$ is sampled and quantized to one bit resolution. Let $r_{m,i}$ denote the $i$th sampling point in the $m$th symbol duration. We have
\begin{equation}
r_{m,i}=\left\{ \begin{array}{ll}
 +1  & \textrm{$r(kT_s+iT)>0$}\\
-1 & \textrm{$r(kT_s+iT)\le 0$}\\
\end{array} \right.
\end{equation}
 We can get the probability 
\begin{equation}
P(r_{m,i}=+1|\bm{c_m})=Q(\sum_{k=0}^{m}c_k p_\text{ref}(m T_s-k T_s + iT))
\end{equation} and
\begin{equation}
P(r_{m,i}=-1|\bm{c_m})=1-Q(\sum_{k=0}^{m}c_k p_\text{ref}(mT_s-kT_s + iT))
\end{equation}
where $Q(x)=1/\sqrt{2\pi}\int_{x}^{+\infty}\exp(-t^2/2) dt$ is the Gaussian $Q$ function.

The digital receiver can get $r_{m,i}$ from the monobit ADC. The main work of the digital processing unit is to take the sampling point $r_{m,i}$ and estimate $\hat{d}_k$ directly.
\begin{figure}
\centering
\includegraphics[width=3.5in]{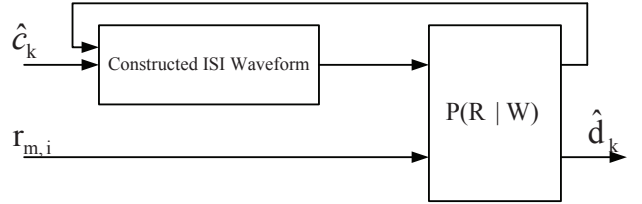}
\caption{Joint decoding diagram}
\label{jointblock}
\end{figure}
\section{Viterbi Decoding and DFE}

\subsection{Joint Viterbi Decoding and DFE}

Joint Viterbi decoding and DFE is a method that estimating the original transmitted information bits $\hat{d_{k}}$. In order to make the BER as small as possible, we make the decision $\hat{d_{k}}$ by its maximum likelihood (ML) probability. Fig. \ref{jointblock} shows the processing of the received points. In \cite{Optimal}, an iterative approach was proposed to estimate channel state information (CSI) by transmitting training symbols. In our research, for simplicity, we assume the full CSI is known in advance. For the convolutional error correction decoder, we use Viterbi algorithm. 

Since the Viterbi algorithm stores information for each state, the complexity of the decoder is proportional to the number of states in the trellis. For a convolutional encoder that has $\mu$ registers, the number of states for this finite state machine (FSM) is $2^\mu$. Define vector $\bm r_{m}=[~r_{m,0},~r_{m,1},~...r_{m,N-1}~]$ and matrix $\bm{R}_k = [~\bm{r}_{0},~\bm{r}_{1},~...\bm{r}_{2k-1}~]$. In this algorithm, we need to construct a reference ISI waveform.
\begin{equation}
p^{k}_\text{wav}(t)=\sum_{m=0}^{2k-1}\hat{c_m} p_\text{ref}(t-mT_s) \nonumber
\end{equation} 
$\hat{c_m}$ is the estimation of $c_m$. Similarly, $w_{m,i} = p_\text{wav}(mT_s+iT)$  denotes the full resolution value of the $i$th point in the $m$th symbol duration. The vector $\bm w_{m}=[~w_{m,0},~w_{m,1},~...w_{m,N-1}~]$ stands for N points in the $m$th duration. The matrix $\bm{W}_k = [~ \bm{w}_{0},~\bm {w}_{1},~...~\bm{w}_{2k-1}~] $ contains the sampling points from the constructed waveform in the duration of $0\sim 2kT_s$. We choose proper $\bm{W}_k$ so that 
\begin{displaymath}
\bm{\hat{d}} = \arg  \max \quad P(\bm{R}_k|\bm{W}_k)
\end{displaymath}

At each step, we update matrix $\bm{W}_k$. At the end of $k$th decoding, $ \hat{c}_{0},~\hat{c}_{1},~...\hat{c}_{2k-1}$ are decided. So for the $(k+1)$th step, the ISI of previous symbols has been estimated. The current signal detection can be viewed as making a decision in AWGN channel. Thanks to the memorylessness of the AWGN channel, we have
\begin{equation}
\log P(\bm{R}_k|\bm{W}_k) = \log \Pi_{k=0}^{2k-1}p(\bm{r}_{k}|\bm{w}_{k})
\end{equation}
Every sampling point is independent with others. Therefore at the $k$th step the log-likelihood probability is 
\begin{equation}
\label{log_pr}
\log P(\bm{R}_k|\bm{W}_k) =\sum_{m=0}^{2k-1} \sum_{i=0}^{N-1} \log p( r_{m,i}| w_{m,i})
\end{equation}
Now we only consider the effect of noise. Then 
\begin{displaymath}
r_{m,i} = \text{sgn} \left(w_{m,i} + n_{m,i}\right)
\end{displaymath}
where $n_{m,i}\sim N(0,1) $. Define $\epsilon_{m,i}=Q(w_{m,i})$, the probability of $r_{m,i}$ can be expressed as follows:

\begin{equation}
\label{probability1}
p( r_{m,i}=+1| w_{m,i}) = 1-\epsilon_{m,i}
\end{equation}
\begin{equation}
\label{probability2}
p( r_{m,i}=-1| w_{m,i})= \epsilon_{m,i}
\end{equation}
Combining (\ref{probability1}) and (\ref{probability2}), we  get
 \begin{equation}
\label{combine_probability}
p( r_{m,i}| c_{m,i}) = 1/2 + r_{m,i}(1/2 - \epsilon_{m,i})
\end{equation}
We substitute (\ref{combine_probability}) into (\ref{log_pr}), the log-likelihood probability is given by:
\begin{equation}
\label{compute}
\log P(\bm{R}_k|\bm{W}_k) =\sum_{m=0}^{2k-1}\sum_{i=0}^{N-1} \log ( 1/2 + r_{m,i}(1/2 - \epsilon_{m,i}) )
\end{equation}

\subsection{Algorithm Description}
The joint Viterbi decoding algorithm attempts to find the maximum log-likelihood probability for each state. It gives the decoding result related to the current probability. As in the standard Viterbi algorithm, this algorithm is based on the trellis. The states and trellis are given by the convolutional encoder. The algorithm needs to store the following information for each state.

1) The surviving path leading to the state.

2) The metric output of this path for the $k$th step. That is, the Viterbi decoder's output for the ``edge" from previous state to current state in the surviving path.

3) The constructed ISI waveform corresponding to the surviving path. This waveform will be used by the next step.

Initialization of the algorithm :

Usually, we start the Viterbi decoding from the all-zero state $s_0$. The metric of $s_0$ is set to be a large positive number.

The algorithm works as follows :

1) Construct a possible ISI waveform for each state. For coding rate $R=1/2$, in the $k$th $(k\ge1)$ step, we get $\bm{r}_{2k-2}$ and $\bm{r}_{2k-1}$ from the received waveform in the duration of $2T_{s}$. We can construct $2^2$ kinds of possible overlapped waveform related to $\{\hat c_{2k-2},\hat c_{2k-1}\}=\{0,0\}$ , $\{0,1\}$ , $\{1,0\}$ or $\{1,1\}$.

2) For each arrived state, there are several leading paths. We compute log-likelihood probability for each path using (11). We also save the constructed waveform of the surviving path.

3) Store the output of the current surviving path and its corresponding code bits. Make a decision of $\hat d_{k}$ for each state. By choosing the surviving path, we make a decision of $\{\hat c_{2k-2},\hat c_{2k-1}\}$ for each state. Then we save $\{\hat c_{2k-2},\hat c_{2k-1}\}$ and $\hat{d}_k$. We also save the constructed waveform for each state.

4) When the last step is finished, only one surviving path is remained. We need to trace back and get the final decoding result $\bm{\hat d}$.

The joint decoding algorithm requires an amount of memory that is proportional to the number of states in the trellis. It also needs memory spaces to store the constructed waveform. The length is $T_\text{max} + T_{s}$.
\subsection{Example}

\begin{figure}
\centering
\includegraphics[width=4in]{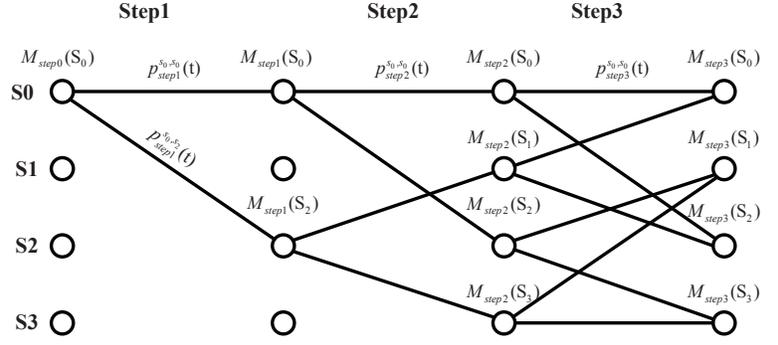}
\caption{Decoding processing in the trellis}
\label{route}
\end{figure}

\begin{figure}
\centering
\includegraphics[width=3in]{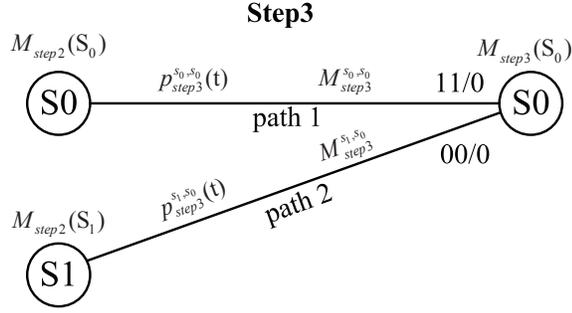}
\caption{Path comparison}
\label{edge}
\end{figure}
Consider a typical trellis with the generator polynomial matrix $g=[5 \quad 7]$ which is shown in Fig. \ref{route}. In the beginning, the metric of $s_0$ is set as 
$M_{step0}(s_0)=I$, where I is a real positive number and $I\gg 0$.

Fig. \ref{route} shows the decoding processing in the trellis. For the first step, we receive $\bm {r}_0$ and $\bm {r}_1$, then our task is construct ``possible" ISI waveforms for the arrived states $s_0$ and $s_2$. There is only one coming path for each of them. We build 
\begin{displaymath}
p_{step1}^{s_0,s_0}(t) = -p_{ref}(t)-p_{ref}(t-T_s) 
\end{displaymath} and
\begin{displaymath}
p_{step1}^{s_0,s_2}(t) = p_{ref}(t)+p_{ref}(t-T_s).
\end{displaymath} $p_{step1}^{s_0,s_2}(t)$ represents the constructed waveform for the path from $s_0$ to $s_2$ in the first step. Let $M_{step1}^{s_0,s_2}$ denotes the path metric from $s_0$ to $s_2$ in the first step. The vectors $\bm {w}_0$ and $\bm {w}_1$ are sampled from $p_{step1}^{s_0,s_0}(t)$ in the duration of $0\sim 2T_s$ and $M_{step1}^{s_0,s_2}$ is computed using (\ref{compute}). We can get $\bm {w}_0$ and $\bm {w}_1$ from $p^{s_0,s_2}_{step1}(t)$ and compute $M^{s_0,s_2}_{step1}$ in the same way. After the first step, we get 
\begin{displaymath}
M_{step1}(s_0)=M_{step0}(s_0)+M^{s_0,s_0}_{step1}
\end{displaymath} and 
\begin{displaymath}
M_{step1}(s_2)=M_{step0}(s_0)+M_{step1}^{s_0,s_2}.
\end{displaymath} We store the ISI waveforms, $\{\hat{c_0},\hat{c_1}\}$, $\hat{d_0}$ and metric for each state. The second step is similar to the first one, let's take the arrived state $s_0$ as an example, we have 
\begin{displaymath}
p_{step2}^{s_0,s_0}(t) =p_{step1}^{s0,s0}(t)-p_{ref}(t-2T_s)-p_{ref}(t-3T_s) 
\end{displaymath}
From the third step, there is a competitive path for each state. In Fig. \ref{edge}, we can easily find that there are two paths leading to $s_0$. Under the assumption that $\{\hat{c_4},\hat{c_5}\}=\{-1,-1\}$, $\hat{d_3}=-1$ for path 1, we build 
\begin{equation}
p_{step3}^{s_0,s_0}(t) = p_{step2}^{s_0,s_0}(t)-p_{ref}(t-4T_s)-p_{ref}(t-5T_s),\nonumber
\end{equation} 
get $\bm {w}_4$ and $\bm {w}_5$ from $p_{step3}^{s_0,s_0}(t)$, and compute $M_{step3}^{s_0,s_0}$. We do the same operation for the second path thus $M_{step3}^{s_1,s_0}$ is achieved. The accumulated path metrics $ M_{step2}(s_0)+M_{step3}^{s_0,s_0}$ and $ M_{step2}(s_1)+M_{step3}^{s_1,s_0}$ should be computed. We pick out the larger one from the two accumulated path metrics, set it as the updated $M_{step3}(s_0)$ and denote its corresponding path as the surviving path. Finally, we store the constructed ISI wave, $\{\hat{c_4},\hat{c_5}\}$, $\hat{d_3}$ and $M_{step3}(s_0)$ for the surviving path. Similarly, for the rest arrived states, we get surviving paths and store the information for each of them. 
When the last step is finished, we pick out the final surviving path which has the largest state metric and complete traceback. Finally, the stored $\hat{d}_k$ in the final surviving path are the decoding result. As is shown in Fig. \ref{route}, if the final surviving path is $s_0 \rightarrow s_2 \rightarrow s_1 \rightarrow s_0$, the result is $\bm{\hat{d}}=\{1,0,0\}$.

\subsection{State Expansion}
\begin{figure}
\centering
\includegraphics[width=2in]{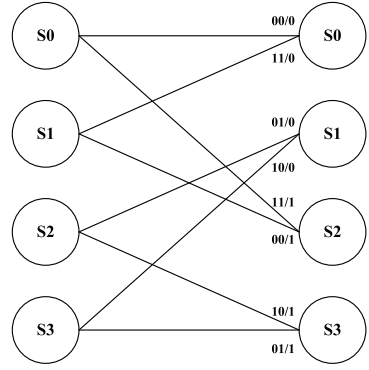}
\caption{4 states}
\label{4states}
\end{figure}
\begin{figure}
\centering
\includegraphics[width=2in]{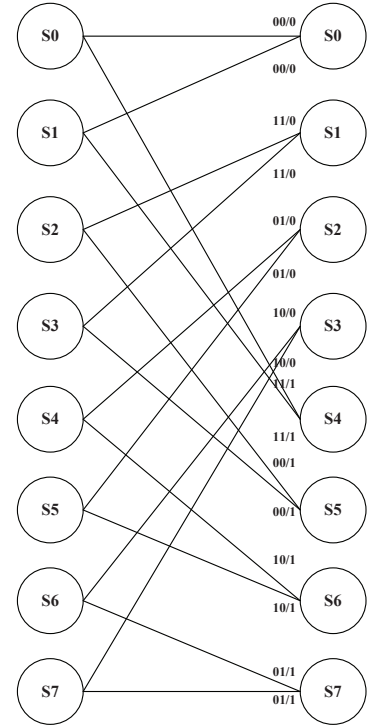}
\caption{8 states after expansion}
\label{8states}
\end{figure}
For a determined trellis, the number of states is given. The BER performance may be poor for lack of states. If we keep more states in the algorithm, we keep more interference situations that may arise. It is desirable to expand states. However, considering the algorithm complexity, we could not expand too many states. The amount of states will change after expansion. Consequently, the trellis also changes.

Fig. \ref{4states} and Fig. \ref{8states} show an example of the original trellis and its expanded trellis respectively. The convolution code rate is $1/2$, generator polynomial matrix $g=[5\quad 7]$.
\section{Simulation Results}

The joint Viterbi decoding and DFE algorithm based on monobit digital receiver are applied to UWB signal processing in this section. Here, the second derivative Gaussian pulse \cite{Gaussian pulse} is used, which can be expressed as follows:
\begin{equation}
p_\text{tr}(t)=(1-4\pi(t/\tau)^2)\exp(-2\pi(t/\tau)^2)
\end{equation}
The constant $\tau$ determines the pulse duration.

The simulation conditions are as follows:

 We use standard CM1 multipath fading channel with 100 realizations \cite{CM1}. The transmitted signal pulse is as (12) with $\tau$=0.22ns. The bandwidth of LPF is $B$ = 5GHz. The sampling period is $T$=0.1ns. Each frame contains $N_{d}=2000$ data. The transmitted symbol rate is 1GHz, that means $T_{s}$=1ns. We have the full CSI. The $T_{max}$ is about 100ns. Fading channel without ISI is also in use. For this type of channel model, we simply let $T_{s}$=100ns. In this simulation system, channel coding is required. We use a typical convolutional code with the rate $R$=1/2. $g = [171\quad 133]$ is the generator polynomial matrix. The SNR is defined as $E_b/N_0 = \sum_{i=1}^{N}{p^2_\text{ref}}(iT)$.

We explain the abbreviation in the simulation results. MB represents the monobit sampling. S represents for soft-decision Viterbi decoding. J represents joint decoding and DFE algorithm. The word CAS indicates that the separate DFE demodulation and Viterbi decoding method is used. The result for fading without ISI channel is denoted by NOISI.

\begin{figure}
\centering
\includegraphics[width=3.5in]{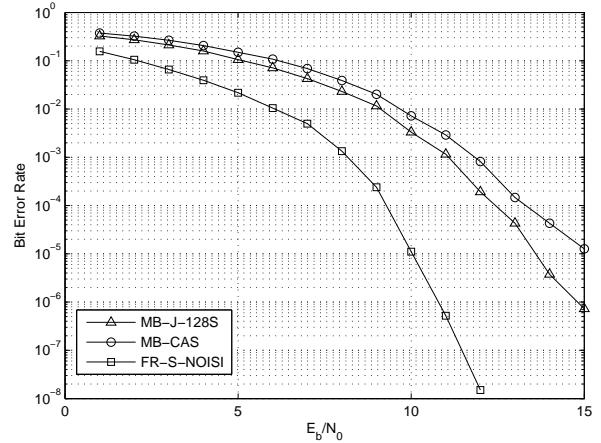}
\caption{Comparison of performance of joint decoding and separate decoding}
\label{CAS}
\end{figure}

Fig. \ref{CAS} shows the BER performance of the joint decoding algorithm and DFE-Viterbi concatenated algorithm in standard CM1 channel.

MB-J-128S: Monobit digital receiver using the joint decoding algorithm that contains 128 states.

MB-CAS: The DFE processing expands its state number to 64 and Viterbi algorithm has 64 states.

FR-S-NOISI: Full resolution detection and soft-decision decoding in CM1 channel without ISI.

\begin{figure}
\centering
\includegraphics[width=3.5in]{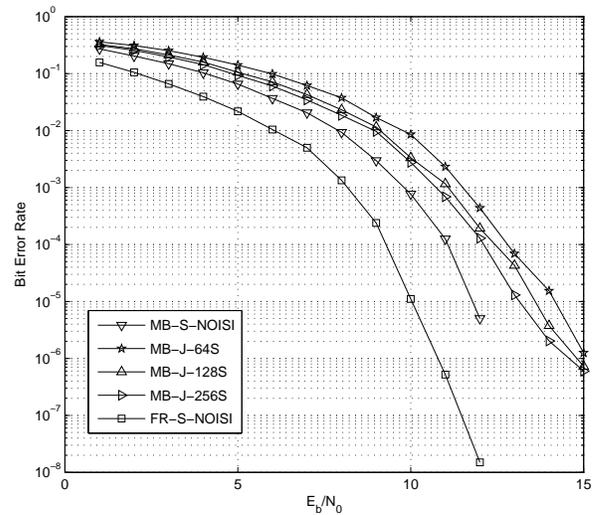}
\caption{BER performance of joint decoding algorithm with different amount of states in fasing isi channel}
\label{diffstates}
\end{figure}

In Fig. \ref{CAS} , we notice that the MB-J-128S has 3dB SNR loss compared to FR-S-NOISI. When SNR is larger than 10dB, We can observe about 1dB gap between the two monobit curves. Fig. \ref{diffstates} shows the effect of state number. It can be seen that the BER decreases as the amount of state increases. The curve MB-S-NOISI is our final destination. As the number of states becomes larger, the gap between BER performance in the ISI channel and that in the channel without ISI becomes smaller. However, when the number of states increases to some extend, there is a BER floor. In the simulation result, state number changes from 128 to 256, the performance does not improve too much. By state expansion, we achieve about 1dB gain. There is still about a 1dB gap between the no ISI BER and joint decoding BER.
\section{Conclusion}

We presented the system model of the monobit digital receivers in ISI channel. The joint Viterbi decoding and DFE algorithm for monobit digital receivers was proposed. We gave the algorithm description and derived the log-likelihood probability of the received signal. State expansion could enhance the BER performance of the algorithm. The BER of the separate decoding method and joint decoding approach was simulated. The simulation shows that the algorithm we proposed has 1dB SNR gain. However, compared to the BER performance of no ISI channel, the algorithm incurs about 1dB loss. There is a 3dB gap between the joint decoding algorithm and FR-S-SEP. This method can be applied to UWB communication systems. It is also an efficient way dealing with ISI for monobit digital receivers. Future research topics include suppressing ISI with estimated CSI by training symbols, increasing the code rate $R$, improving the BER performance.





%

\end{document}